\documentclass[prl,aps,preprintnumbers,superscriptaddress,twocolumn]{revtex4}
\usepackage{epsfig,amsmath,amssymb,amsfonts}

\newcommand{\ket}[1]{\left | \, #1 \right\rangle}

\newcommand{\half}{\mbox{$\textstyle \frac{1}{2}$}}
\def\opone{\leavevmode\hbox{\small1\kern-3.8pt\normalsize1}}

\newcommand{\vis}{\text{v}}

\begin{document}
\title{Surface Acoustic Wave Single-Electron Interferometry}
    \author{Roberta \surname{Rodriquez}} \email{rr269@damtp.cam.ac.uk}
    \affiliation{Centre for Quantum Computation, DAMTP,
    University of Cambridge, Wilberforce Road, Cambridge CB3 0WA, United Kingdom}
    \affiliation{Cavendish Laboratory,
    University of Cambridge, Madingley Road, Cambridge CB3 0HE, United Kingdom}
    \author{Daniel K. L. \surname{Oi}}
    \affiliation{Centre for Quantum Computation, DAMTP,
    University of Cambridge, Wilberforce Road, Cambridge CB3 0WA, United Kingdom}
    \author{Crispin H. W. Barnes}
    \affiliation{Cavendish Laboratory,
    University of Cambridge, Madingley Road, Cambridge CB3 0HE, United Kingdom}
    \author{Masaya Kataoka}
    \affiliation{Cavendish Laboratory,
    University of Cambridge, Madingley Road, Cambridge CB3 0HE, United Kingdom}
    \author{Toshio \surname{Ohshima}}
    \affiliation{Centre for Quantum Computation, DAMTP, University of
      Cambridge, Wilberforce Road, Cambridge CB3 0WA, United Kingdom}
    \affiliation{Fujitsu Laboratories of Europe Ltd. (FLE)
Hayes Park Central, Hayes End Road, Hayes, Middlesex UB4 8FE,
United Kingdom}
    \author{Artur K. \surname{Ekert}}
    \affiliation{Centre for Quantum Computation, DAMTP,
    University of Cambridge, Wilberforce Road, Cambridge CB3 0WA, United Kingdom}


\begin{abstract}
  We propose an experiment to observe interference of a single electron as it
  is transported along two parallel quasi-one-dimensional channels trapped in
  a single minimum of a travelling periodic electric field.  The experimental
  device is a modification of the surface acoustic wave (SAW) based quantum
  processor. Interference is achieved by creating a superposition of spatial
  wavefunctions between the two channels and inducing a relative phase shift
  via either a transverse electric field or a magnetic field. The interference
  can be used to estimate the decoherence time of an electron in this type of
  solid-state device.
\end{abstract}

\maketitle \preprint{Cambridge, \today}


Constructing a solid-state single-electron interferometer poses
many challenges, especially single-electron transport through the
device. Recent experiments on electron
interferometers~\cite{JCSHMS2003,WNDFEH2003} and double quantum
dots~\cite{fuji} have demonstrated interference, but do not deal
with single electrons. These experiments have to take into account
many-particle effects, the behaviour of electrons as
quasi-particles, and the validity of the application of theories
such as Fermi liquid theory. Besides not showing true single
particle interference, these factors obscure the fundamental
electron coherence time, which is of crucial importance for many
prospective solid state quantum information processing
schemes~\cite{CLSZ95,Unruh,Shor,Ekert,Steane}.

Electron quantization using surface acoustic waves (SAW),
originally studied in the context of current
standards~\cite{STPRFFSJ1996,TSPSFLRJ1997}, has recently lead to a
proposal for the implementation of a quantum processor in the
solid-state that uses this mechanism~\cite{BarShiRob00}.
Advantages of the proposed SAW devices include the unique feature
of creating a completely polarised initial state and of making
ensemble measurements over billions of identical computations.
Additionally, these systems are similar to quantum dots, but have
the advantage that manipulation of qubits can be done with static
potentials on surface gates without the need for expensive
high-frequency pulse generation~\cite{fuji}. Furthermore, the
mechanism of SAW transport eliminates the problem of
backscattering from discontinuities in the electron trajectory
which also detracts from the ideal interferometry
experiment~\footnote{Numerical studies by~\cite{Bertoni} suggest
that SAW assisted transport increases quantum coherence over
ballistic transport.}~\footnote{As this paper was being completed
we were made aware of the work of~\cite{Bertoni} who have
considered a similar situation.}. This opens up the range of
mechanisms for inducing relative phase shifts required to observe
interference fringes.

The acoustoelectric devices we consider in this paper are
fabricated on modulation doped GaAs-AlGaAs heterostructures.
Because GaAs is a piezoelectric material, applying a
radio-frequency potential difference between a pair of
interdigitated transducers produces vibrations that propagate
through the structure as longitudinal waves (SAWs), which in turn
induce an electrostatic potential. The SAWs then travel across the
2-dimensional electron gas and through a mesa patterned with
surface gates that define two parallel quasi-one-dimensional
channels. By altering the static potential on the surface gates it
is possible to trap a single electron in each SAW potential
minimum in each of the two channels with an accuracy greater than
1 part in $10^5$~\cite{cunningham}. A two level quantum system
(qubit) can be defined by the presence of a single electron in
either the lower or the upper channel ($\ket{0}$ and $\ket{1}$
respectively). Single qubit rotations can be implemented by
variations in the static potentials defined by surface gates. The
probability of the presence of an electron in either channel can
be measured directly from the current output of each channel via
Ohmic contacts.

A Mach-Zender single particle interferometer can be constructed
from a single qubit SAW processor by a combination of $\sigma_x$
and $\sigma_z$ gates. The size of the interference fringes gives
an indication of the fidelity of device which is a combination of
the individual gate fidelities and decoherence. By varying the
effective length of the interferometer, the dephasing time of
single electrons in this system can be estimated, which is
expected to be the limiting factor for coherent manipulation of
these systems.


Decoherence of qubit can be characterised by two timescales, the
$T_1$ and the $T_2$ time, which are a measure of the rate at which
the system experiences unwanted transitions and dephasing between
quantum levels respectively. In the Bloch sphere
picture~\cite{BLOCH,Nielsen,OI2001}, the $T_1$ (amplitude damping)
time is associated with the contraction of the Bloch sphere along
the z-axis, in conjunction with a symmetrical contraction along
the x- and y-axes consonant with complete positivity~\footnote{The
squeezing of the Bloch sphere in orthogonal directions
  are constrained by the structure of quantum mechanics to obey
  $|\eta_x\pm\eta_y|\ge|1\pm\eta_z|$~\cite{OI2001}. This constraint stems from
  linearity and the possibility of the system being entangled with other
  systems~\cite{kraus}.}. This transforms a pure state to a completely mixed
state.  The $T_2$ (phase relaxation) time is associated with the
contraction of the x- and y-axes only, resulting in as shrinkage
of the Bloch sphere to a line along the z-axis. In the Markovian
regime, an initially pure state,
$\ket{\psi}=\alpha\ket{0}+\beta\ket{1}$, evolves under phase
relaxation as
\begin{equation}\label{eq:decoh}
\rho_s(t) = \left(
 \begin{array}{cc}
 |\alpha|^2 & \alpha \beta^* e^{-t/T_2}\\
 \alpha^* \beta e^{-t/T_2} & |\beta|^2 \\
 \end{array}
\right).
\end{equation}
The off-diagonal terms (coherences), responsible for interference, decrease in
magnitude exponentially, where $T_2$ is the $1/e$ time constant.


\begin{figure}[!htp]
\begin{center}
\includegraphics[width=0.38\textwidth]{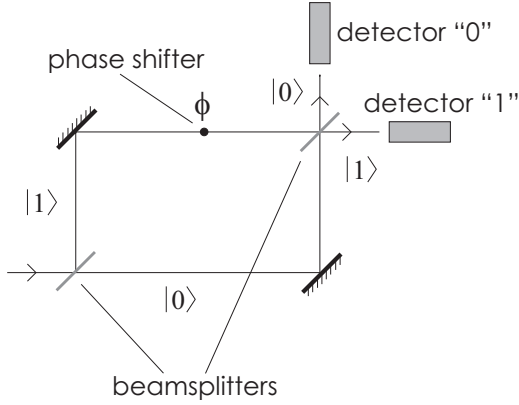}
\end{center}
\caption{A Mach Zender interferometer. A single particle at a time is sent
  horizontally towards the first beamsplitter. We label the state of the
  particle in the upper and lower arms of the interferometer $\ket{0}$ and
  $\ket{1}$ respectively. A phase shift is introduced into the upper arm. The
  two paths are directed to interfere at a second beamsplitter. Particle
  detectors determine from which direction the particle exits the interferometer.}
\label{fig:mach}
\end{figure}

\begin{figure}[!htp]
\includegraphics[width=0.45\textwidth]{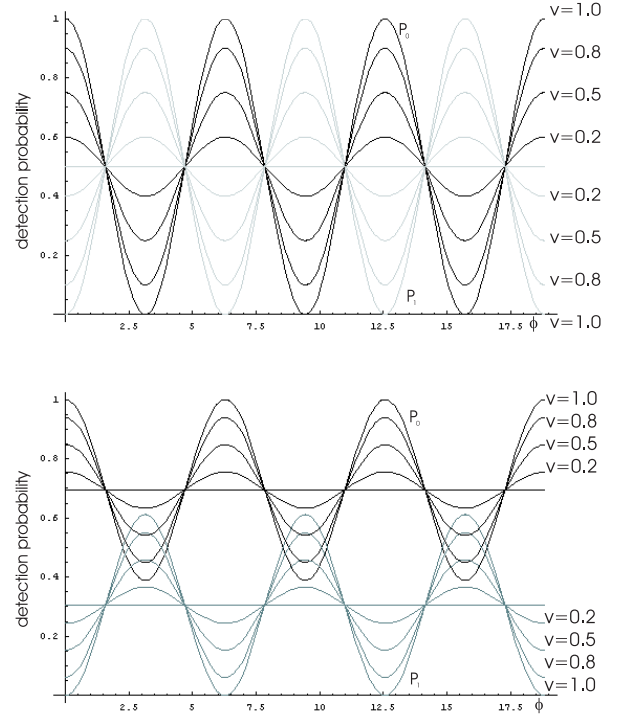}
\caption{\small{Interference patterns showing reduction in
    visibility as decoherence increases, as well as divergence of the two
    detector curves if the beamsplitter is not 50:50. The upper figure is for
    $\theta=\pi/4$, the lower one for $\theta=\pi/8$.}}\label{fig:apattern}
\end{figure}

A Mach-Zender interferometer is shown in Fig.~\ref{fig:mach}.
Initially, a particle is in the localised state $\ket{0}$
travelling horizontally towards the first beamsplitter. The
actions of the beamsplitters, having transmittances
$t=\cos^2\theta$ and reflectances $r=\sin^2\theta$, and phase
shifter can be expressed as unitary operations, $U_{BS}= \left(
 \begin{array}{cc}
 \cos \theta &   -\sin \theta \\
 e^{i\gamma}\sin \theta &  e^{i\gamma} \cos \theta  \\
 \end{array}
\right) $ and $\varphi= \left(\begin{array}{cc}
    1 & 0 \\
    0 & e^{i\phi}\\
 \end{array}
\right)$ respectively. The state may experience dephasing for a period of
$\tau$, the transit time between the two beamsplitters. The final state after
the second beamsplitter is
\begin{eqnarray}
\rho_{00}&=&\cos^4\theta+\sin^4\theta +\half \vis \sin^2 2\theta \cos(\gamma+\phi)\nonumber\\
\rho_{01}=\rho_{10}^{*}&=& \half e^{-i\gamma}\sin 2\theta ( \cos 2\theta +\vis e^{i(\gamma+\phi)}  \nonumber\\
& & - 2\vis \cos^2 \theta \cos (\gamma+\phi))  \nonumber \\
\rho_{11}&=&\half\ \sin^2 2\theta
\left(1-\vis\cos(\gamma+\phi)\right),\nonumber
\end{eqnarray}
where $\vis=e^{-\tau/T_2}$. The probabilities of each detector
clicking therefore are
\begin{subequations}
\begin{eqnarray}
P_0 &=&\cos^4\theta+\sin^4\theta +\half \vis \sin^2 2\theta
\cos(\gamma+\phi)
\\
P_1&=&\half\sin^2 2\theta(1-\vis\cos(\gamma+\phi)).
\end{eqnarray}
\end{subequations}
By varying $\phi$, interference fringes can be observed
(Fig.\ref{fig:apattern}). Using the standard definition of
visibility, $\;\;\;$
$\vis=\frac{P_{max}-P_{min}}{P_{max}+P_{min}}\label{eq:vis}$, we
find that
\begin{subequations}
\begin{eqnarray}
\vis_0&=&\frac{ \vis \sin^2 2 \theta}{2(\cos^4 \theta +
\sin^4 \theta)} \\
\vis_1&=& \vis, \quad \forall\  \theta.
\label{eq:z}
\end{eqnarray}
\end{subequations}
Therefore $\vis_1$ only depends on the dephasing.

If the beamsplitters have different splitting ratios, the
interference pattern will depend on $\vis$ and the two angles
$\theta_1$ and $\theta_2$. The average of $P_1$ or $P_2$ (with
respect to $\phi$) will be $\frac{1}{2}$ if at least one of the
beamsplitter ratios is $50:50$. This allows the possibility of
tuning the interferometer by adjusting the first beamsplitter
until the average value of $P_0$ or $P_1$ is $\frac{1}{2}$, and
then adjusting the second beamsplitter to maximize the visibility.


\begin{figure}[!htp]
\includegraphics[width=0.45\textwidth]{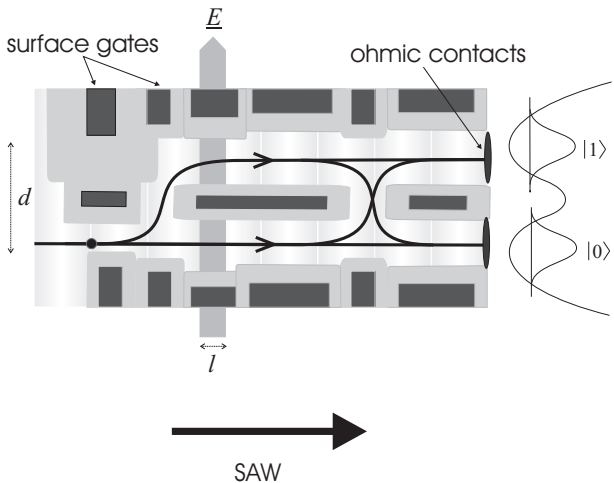}
\caption{\label{fig:interf}Single-electron interferometer.
  A surface acoustic wave propagates from left to right. Single electrons are
  transported by
   the SAW along 1-D channels defined by surface electrodes
  parallel to the direction of SAW propagation.  By lowering the potential
  between two channels (by a suitable gap in the surface electrodes), an
  electron may coherently tunnel laterally like in a beamsplitter. Biasing the
  channels relative to each other induces a phase gate.}
\end{figure}

A two-channel SAW device is shown in Fig.~\ref{fig:interf}. One channel is
blocked off so that only one electron is carried in the wavefront of each SAW
potential minimum. Information is encoded on the position of the electron, so
that localisation to the upper and lower channels corresponds to the qubit
states $\ket{0}$ and $\ket{1}$ respectively. A superposition of the two states
can be created by lowering the potential barrier between the two channels with
the aid of a gap in the surface gates.  While the electron is in the region of
the gap, it can tunnel between the two channels. Its dynamics can be described
by the effective Hamiltonian
\begin{equation}
H=\frac{1}{2} \epsilon \sigma_z +\frac{1}{2} \Delta \sigma_x
\end{equation}
where $\sigma_x$ and $\sigma_z$  are the Pauli matrices acting on
$\ket{0}$ and $\ket{1}$, and $\epsilon$ is the energy splitting
between the localised electron energy levels in each well. For
small $\epsilon$ and for $\ket{\psi(t=0)}=\ket{0}$,
\begin{equation}\label{eq:tunnel}
\ket{\psi(t)}=\cos(\alpha t)\ket{0}-i\sin(\alpha t)\ket{1},
\end{equation}
where $\alpha=\Delta/\hbar$, and $\Delta$ is the tunnelling
frequency. The tunnelling time is determined by the size of the
tunnelling region, since the velocity of the SAW is fixed, so that
Eq.~(\ref{eq:tunnel}) describes the map $\ket{0}\mapsto
\cos\theta\ket{0}-i\sin\theta\ket{1}$, where $\theta$ is now
related to the size of the barrier. The tunnelling region
therefore acts like a beamsplitter.

In order to observe single-electron interference, we introduce a
relative phase shift $\phi$ between the two paths which can be
achieved in several ways.  One can induce an asymmetry in the
double well potential by means of a transverse electric field, as
shown in Fig.~\ref{fig:interf}, or by narrowing the 1D-channel
confinement potential. Alternatively, one could employ the
Aharonov-Bohm effect, which has already been observed in
GaAs/AlGaAs heterostructure devices~\cite{TCCCMBH87}.

Introducing an asymmetry to the double well potential via a transverse electric field
separates the eigenstates of the systems into localised single particle
eigenfunctions, evolving with different energies:
\begin{equation}
\ket{\psi}=\cos\theta e^{-iE_0t/\hbar}\ket{0}- i\sin\theta
e^{-iE_1t/\hbar}\ket{1}.
\end{equation}
The relative phase difference between the two paths is therefore
given by the energy difference $\epsilon=E_0-E_1$ between the two
localised states
\begin{equation}\label{eq:phase2} \Delta \phi=
\epsilon= \frac{e}{\hbar}  \int  V dt,
\end{equation}
where $V$ is the voltage difference between the two channels and
$e$ is the electronic charge. Since the electrons are transported
by the SAW, $\int dt=\tau= l/v$ where $l$ is the length of the
channel region experiencing the electric field and $v$ is the
velocity of the SAW ($\sim 2700$m/s in GaAs). We can then rewrite
Eq.~(\ref{eq:phase2}) as
\begin{equation}\label{eq:phase3}
\Delta\phi=\frac{e|\vec{E}|d }{\hbar}\frac{l}{v},
\end{equation}
since $V=\vec{E}.\vec{d}$, where $\vec{E}$ is the the electric
field and $\vec{d}$ is the displacement between the two channels,
and therefore explicitly calculate $\Delta\phi$.

The lowest electron temperature achievable in a $^{3}$He -
$^{4}$He dilution refrigerator is realistically around $100mK$
($\sim 10\mu eV$), assuming that microwave heating is minimized.
We take this thermal energy as the resolution of the experiment.
In order to obtain clearly defined oscillations, the minimum
transverse potential change needed for each $2\pi$ phase change is
$\sim 100\mu$V, corresponding to a maximum phase gate length of
$0.1\mu m$. We cannot have a longer gate without decreasing the
number of readings per fringe, given the voltage resolution due to
thermal noise. We also require observation of several periods in
order to obtain a good estimate of the visibility.

If the relative phase shift is introduced via the Aharonov-Bohm
effect~\cite{AB1959}, we have that
\begin{equation}
\Delta \phi= \frac{e}{\hbar} \int \vec{B} \cdot \vec{n} dS
\end{equation}
where $S$ is the surface enclosed by the two paths of the
interferometer. In our setup, in order to obtain a $2\pi$ phase
shift, if the area enclosed by two paths is of the order of $\sim
0.2\mu m^2$, a $|\vec{B}|$ field change of the order of $\sim
20mT$ is required. Interference of electrons has already been
observed in the presence of large magnetic fields
in~\cite{JCSHMS2003}; we thus expect that this small magnetic
field should not produce much additional decoherence.

\begin{figure}[!htp]
\includegraphics[width=0.45\textwidth]{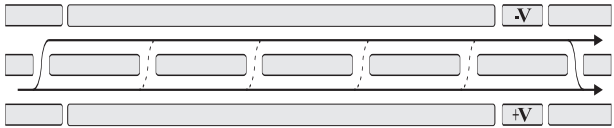}
\caption{Configuration for measuring the decoherence rate. By
  varying the central gate bias, the gaps in the central barrier are opened and
  closed in such a manner that the path length can be varied, hence varying the
  time the electron is in superposition between the upper and lower
  channels.}\label{fig:varlength}
\end{figure}

To measure the dephasing rate, we need to subject the
superposition of localised electron states to increasing lengths
of time and measure for each length the reduction of the
visibility. This can be achieved by lengthening the effective path
length of the interferometer, as shown in
Fig.~\ref{fig:varlength}. We require at least 5 different times to
obtain a reasonable estimate of $T_2$. The longest interferometer
transit time should be of the order of $2.3 \times T_2$, if we
require the minimum visibility to be 10\% of the initial
visibility. Although absolute estimates the $T_2$ time do not
exist, recent experiments place a lower bound on decoherence of
$\sim 1ns$~\cite{fuji}. Using this value, we find that the longest
channel setting needs to be of the order of $v\tau\sim 6\mu m$.
Increments in channel distance between each setting thus need to
be of $\sim 1.2\mu m$ or less. This is easily achievable using
current electron-beam lithography technology.


The $T_1$ time, corresponding to unwanted tunnelling, can be made extremely
long in between the two beamsplitter regions and may be ignored.  In the
tunnelling regions however, effects like scattering from fluctuating impurity
potentials (random telegraph noise) do become important.  Estimates of the
decoherence time for similar tunnelling regions have been made for a double dot
system and found to be at least $1ns$~\cite{fuji}.  Since our tunnel regions
are $\sim 300nm$ long, the electron traverses them in a less than $100ps$, so
we expect these errors to be small. In any case, these gate errors are constant
and thus one can factor out their effect to determine $T_2$.  Since the
electron transported by the SAW is shielded from many particle effects, our
system may show higher coherence than multi-electron quantum dots~\cite{fuji}.

Increasing the channel length to estimate the dephasing time will
be a challenge. A main concern will be that the environment of the
qubits will change. However, the increase in static impurities
will be small (for an average impurity density of $\sim 1\mu
m^{-1}$) and techniques exists to `delete' their effects on the
qubits, once their presence is located~\cite{crook}. Calibration
of the beamsplitters is vital to eliminate the contribution of
mismatched splitting ratios to the variation in interference
visibility.

We do not include in our analysis decoherence arising from
spin-orbit coupling. This, however, we expect to be negligible
because of the much longer decoherence times supported by the spin
degree of freedom~\cite{KA1998}.

Finally, this device can also be used as an electric field
measuring device, since changes in the transverse electric field
will result in changes in the interference pattern. By means of a
feedback circuit, the absolute size of the field can be measured.
This measurement will be subject to shot noise,
$\sqrt{N}/N=1\sqrt{N}$, where $N=f \Delta t$ is the total number
of electrons collected in time $\Delta t$ with  SAWs of frequency
$f$. There is a trade-off between increased sensitivity, by using
a longer $\Delta t$, and measurement bandwith.

We would like to thank the Schiff Foundation, Fujitsu, EU projects
RESQ (IST-2001-37559) and TOPQIP (IST-2001-39215), Sidney Sussex
College and the CMI collaboration for financial support, and
Valery Talyanskii for stimulating conversations.


\begin{thebibliography}{21}
\expandafter\ifx\csname
natexlab\endcsname\relax\def\natexlab#1{#1}\fi
\expandafter\ifx\csname bibnamefont\endcsname\relax
  \def\bibnamefont#1{#1}\fi
\expandafter\ifx\csname bibfnamefont\endcsname\relax
  \def\bibfnamefont#1{#1}\fi
\expandafter\ifx\csname citenamefont\endcsname\relax
  \def\citenamefont#1{#1}\fi
\expandafter\ifx\csname url\endcsname\relax
  \def\url#1{\texttt{#1}}\fi
\expandafter\ifx\csname
urlprefix\endcsname\relax\def\urlprefix{URL }\fi
\providecommand{\bibinfo}[2]{#2}
\providecommand{\eprint}[2][]{\url{#2}}

\bibitem[{\citenamefont{Ji et~al.}(2003)\citenamefont{Ji, CHung, Heiblum,
  Mahalu, and Shtrikman}}]{JCSHMS2003}
\bibinfo{author}{\bibfnamefont{Y.}~\bibnamefont{Ji}},
  \bibinfo{author}{\bibfnamefont{Y.}~\bibnamefont{CHung}},
  \bibinfo{author}{\bibfnamefont{M.}~\bibnamefont{Heiblum}},
  \bibinfo{author}{\bibfnamefont{D.}~\bibnamefont{Mahalu}}, \bibnamefont{and}
  \bibinfo{author}{\bibfnamefont{H.}~\bibnamefont{Shtrikman}},
  \bibinfo{journal}{Nature} \textbf{\bibinfo{volume}{422}},
  \bibinfo{pages}{415} (\bibinfo{year}{2003}).

\bibitem[{\citenamefont{van~der Wiel et~al.}(2003)\citenamefont{van~der Wiel,
  Y.Nazarov, DeFranceschi, Fujisawa, Elzerman, and Huilzeling}}]{WNDFEH2003}
\bibinfo{author}{\bibfnamefont{W.}~\bibnamefont{van~der Wiel}},
  \bibinfo{author}{\bibnamefont{Y.Nazarov}},
  \bibinfo{author}{\bibfnamefont{S.}~\bibnamefont{DeFranceschi}},
  \bibinfo{author}{\bibfnamefont{T.}~\bibnamefont{Fujisawa}},
  \bibinfo{author}{\bibfnamefont{J.}~\bibnamefont{Elzerman}}, \bibnamefont{and}
  \bibinfo{author}{\bibfnamefont{E.}~\bibnamefont{Huilzeling}},
  \bibinfo{journal}{Physical Review B} \textbf{\bibinfo{volume}{67}},
  \bibinfo{pages}{033307} (\bibinfo{year}{2003}).

\bibitem[{\citenamefont{Hayashi et~al.}(2003)\citenamefont{Hayashi, Fujisawa,
  Cheong, Jeong, and Hirayama}}]{fuji}
\bibinfo{author}{\bibfnamefont{T.}~\bibnamefont{Hayashi}},
  \bibinfo{author}{\bibfnamefont{T.}~\bibnamefont{Fujisawa}},
  \bibinfo{author}{\bibfnamefont{H.~D.} \bibnamefont{Cheong}},
  \bibinfo{author}{\bibfnamefont{Y.~H.} \bibnamefont{Jeong}}, \bibnamefont{and}
  \bibinfo{author}{\bibfnamefont{Y.}~\bibnamefont{Hirayama}},
  \bibinfo{journal}{Phys. Rev. Lett.} \textbf{\bibinfo{volume}{91}},
  \bibinfo{pages}{226804} (\bibinfo{year}{2003}).

\bibitem[{\citenamefont{Chuang et~al.}(1995)\citenamefont{Chuang, Laflamme,
  Shor, and Zurek}}]{CLSZ95}
\bibinfo{author}{\bibfnamefont{I.~L.} \bibnamefont{Chuang}},
  \bibinfo{author}{\bibfnamefont{R.}~\bibnamefont{Laflamme}},
  \bibinfo{author}{\bibfnamefont{P.~W.} \bibnamefont{Shor}}, \bibnamefont{and}
  \bibinfo{author}{\bibfnamefont{W.~H.} \bibnamefont{Zurek}},
  \bibinfo{journal}{Science} \textbf{\bibinfo{volume}{270}},
  \bibinfo{pages}{1633} (\bibinfo{year}{1995}).

\bibitem[{\citenamefont{Unruh}(1995)}]{Unruh}
\bibinfo{author}{\bibfnamefont{W.~G.} \bibnamefont{Unruh}},
  \bibinfo{journal}{Phys. Rev. A} \textbf{\bibinfo{volume}{51}},
  \bibinfo{pages}{992} (\bibinfo{year}{1995}).

\bibitem[{\citenamefont{Shor}(1995)}]{Shor}
\bibinfo{author}{\bibfnamefont{P.}~\bibnamefont{Shor}}, \bibinfo{journal}{Phys.
  Rev. A} \textbf{\bibinfo{volume}{52}}, \bibinfo{pages}{2493}
  (\bibinfo{year}{1995}).

\bibitem[{\citenamefont{Ekert and Macchiavello}(1996)}]{Ekert}
\bibinfo{author}{\bibfnamefont{A.}~\bibnamefont{Ekert}} \bibnamefont{and}
  \bibinfo{author}{\bibfnamefont{C.}~\bibnamefont{Macchiavello}},
  \bibinfo{journal}{Phys. Rev. Lett.} \textbf{\bibinfo{volume}{77}},
  \bibinfo{pages}{2585} (\bibinfo{year}{1996}).

\bibitem[{\citenamefont{Steane}(2003)}]{Steane}
\bibinfo{author}{\bibfnamefont{A.}~\bibnamefont{Steane}},
  \bibinfo{journal}{Phys. Rev. A} \textbf{\bibinfo{volume}{68}},
  \bibinfo{pages}{042322} (\bibinfo{year}{2003}).

\bibitem[{\citenamefont{Shilton et~al.}(1996)\citenamefont{Shilton, Talyanskii,
  Pepper, Ritchie, Frost, Ford, Smith, and Jones}}]{STPRFFSJ1996}
\bibinfo{author}{\bibfnamefont{J.~M.} \bibnamefont{Shilton}},
  \bibinfo{author}{\bibfnamefont{V.~I.} \bibnamefont{Talyanskii}},
  \bibinfo{author}{\bibfnamefont{M.}~\bibnamefont{Pepper}},
  \bibinfo{author}{\bibfnamefont{D.~A.} \bibnamefont{Ritchie}},
  \bibinfo{author}{\bibfnamefont{J.~E.~F.} \bibnamefont{Frost}},
  \bibinfo{author}{\bibfnamefont{C.~J.~B.} \bibnamefont{Ford}},
  \bibinfo{author}{\bibfnamefont{C.~G.} \bibnamefont{Smith}}, \bibnamefont{and}
  \bibinfo{author}{\bibfnamefont{G.~A.~C.} \bibnamefont{Jones}},
  \bibinfo{journal}{J. Phys.: Cond. Mat.} \textbf{\bibinfo{volume}{8}},
  \bibinfo{pages}{531} (\bibinfo{year}{1996}).

\bibitem[{\citenamefont{Talyanskii et~al.}(1997)\citenamefont{Talyanskii,
  Shilton, Pepper, Smith, Ford, Linfield, Ritchie, and Jones}}]{TSPSFLRJ1997}
\bibinfo{author}{\bibfnamefont{V.~I.} \bibnamefont{Talyanskii}},
  \bibinfo{author}{\bibfnamefont{J.~M.} \bibnamefont{Shilton}},
  \bibinfo{author}{\bibfnamefont{M.}~\bibnamefont{Pepper}},
  \bibinfo{author}{\bibfnamefont{C.~G.} \bibnamefont{Smith}},
  \bibinfo{author}{\bibfnamefont{C.~J.~B.} \bibnamefont{Ford}},
  \bibinfo{author}{\bibfnamefont{E.~H.} \bibnamefont{Linfield}},
  \bibinfo{author}{\bibfnamefont{D.~A.} \bibnamefont{Ritchie}},
  \bibnamefont{and} \bibinfo{author}{\bibfnamefont{G.~A.~C.}
  \bibnamefont{Jones}}, \bibinfo{journal}{Phys. Rev. B}
  \textbf{\bibinfo{volume}{56}}, \bibinfo{pages}{15180} (\bibinfo{year}{1997}).

\bibitem[{\citenamefont{Barnes et~al.}(2000)\citenamefont{Barnes, Shilton, and
  Robinson}}]{BarShiRob00}
\bibinfo{author}{\bibfnamefont{C.~H.} \bibnamefont{Barnes}},
  \bibinfo{author}{\bibfnamefont{J.~M.} \bibnamefont{Shilton}},
  \bibnamefont{and} \bibinfo{author}{\bibfnamefont{A.~M.}
  \bibnamefont{Robinson}}, \bibinfo{journal}{Physical Review B}
  \textbf{\bibinfo{volume}{62}}, \bibinfo{pages}{8410} (\bibinfo{year}{2000}).

\bibitem[{\citenamefont{Cunningham et~al.}(2000)\citenamefont{Cunningham,
  Talyanskii, Shilton, Pepper, Kristensen, and Lindelof}}]{cunningham}
\bibinfo{author}{\bibfnamefont{J.}~\bibnamefont{Cunningham}},
  \bibinfo{author}{\bibfnamefont{V.~I.} \bibnamefont{Talyanskii}},
  \bibinfo{author}{\bibfnamefont{J.~M.} \bibnamefont{Shilton}},
  \bibinfo{author}{\bibfnamefont{M.}~\bibnamefont{Pepper}},
  \bibinfo{author}{\bibfnamefont{A.}~\bibnamefont{Kristensen}},
  \bibnamefont{and} \bibinfo{author}{\bibfnamefont{P.~E.}
  \bibnamefont{Lindelof}}, \bibinfo{journal}{Physica B}
  \textbf{\bibinfo{volume}{280}}, \bibinfo{pages}{493} (\bibinfo{year}{2000}).

\bibitem[{\citenamefont{Bloch}(1946)}]{BLOCH}
\bibinfo{author}{\bibfnamefont{F.}~\bibnamefont{Bloch}},
  \bibinfo{journal}{Phys. Rev.} \textbf{\bibinfo{volume}{70}},
  \bibinfo{pages}{460} (\bibinfo{year}{1946}).

\bibitem[{\citenamefont{Nielsen and Chuang}(2000)}]{Nielsen}
\bibinfo{author}{\bibfnamefont{M.}~\bibnamefont{Nielsen}} \bibnamefont{and}
  \bibinfo{author}{\bibfnamefont{I.}~\bibnamefont{Chuang}},
  \emph{\bibinfo{title}{Quantum Computation and Quantum Information}}
  (\bibinfo{publisher}{Cambridge}, \bibinfo{year}{2000}).

\bibitem[{\citenamefont{Oi}(2001)}]{OI2001}
\bibinfo{author}{\bibfnamefont{D.~K.~L.} \bibnamefont{Oi}},
  \emph{\bibinfo{title}{The geometry of single qubit maps}},
  \bibinfo{howpublished}{e-print quant-ph/0106035} (\bibinfo{year}{2001}).

\bibitem[{\citenamefont{G.Timp et~al.}(1987)\citenamefont{G.Timp, A.M.Chang,
  J.E.Cunningham, T.Y.Chang, P.Mankiewich, R.Behringer, and
  R.E.Howard}}]{TCCCMBH87}
\bibinfo{author}{\bibnamefont{G.Timp}},
  \bibinfo{author}{\bibnamefont{A.M.Chang}},
  \bibinfo{author}{\bibnamefont{J.E.Cunningham}},
  \bibinfo{author}{\bibnamefont{T.Y.Chang}},
  \bibinfo{author}{\bibnamefont{P.Mankiewich}},
  \bibinfo{author}{\bibnamefont{R.Behringer}}, \bibnamefont{and}
  \bibinfo{author}{\bibnamefont{R.E.Howard}}, \bibinfo{journal}{Phys. Rev.
  Lett.} \textbf{\bibinfo{volume}{58}}, \bibinfo{pages}{2814}
  (\bibinfo{year}{1987}).

\bibitem[{\citenamefont{Aharonov and Bohm}(1959)}]{AB1959}
\bibinfo{author}{\bibfnamefont{Y.}~\bibnamefont{Aharonov}} \bibnamefont{and}
  \bibinfo{author}{\bibfnamefont{D.}~\bibnamefont{Bohm}},
  \bibinfo{journal}{Phys. Rev.} \textbf{\bibinfo{volume}{115}},
  \bibinfo{pages}{485} (\bibinfo{year}{1959}).

\bibitem[{\citenamefont{Crook et~al.}(2003)\citenamefont{Crook, Graham, Smith,
  Farrer, Beere, and Ritchie}}]{crook}
\bibinfo{author}{\bibfnamefont{R.}~\bibnamefont{Crook}},
  \bibinfo{author}{\bibfnamefont{A.}~\bibnamefont{Graham}},
  \bibinfo{author}{\bibfnamefont{C.}~\bibnamefont{Smith}},
  \bibinfo{author}{\bibfnamefont{I.}~\bibnamefont{Farrer}},
  \bibinfo{author}{\bibfnamefont{H.}~\bibnamefont{Beere}}, \bibnamefont{and}
  \bibinfo{author}{\bibfnamefont{D.}~\bibnamefont{Ritchie}},
  \bibinfo{journal}{Nature} \textbf{\bibinfo{volume}{424}},
  \bibinfo{pages}{751} (\bibinfo{year}{2003}).

\bibitem[{\citenamefont{Kikkawa and Awschalom}(1998)}]{KA1998}
\bibinfo{author}{\bibfnamefont{J.~M.} \bibnamefont{Kikkawa}} \bibnamefont{and}
  \bibinfo{author}{\bibfnamefont{D.~D.} \bibnamefont{Awschalom}},
  \bibinfo{journal}{Phys. Rev. Lett.} \textbf{\bibinfo{volume}{80}},
  \bibinfo{pages}{4313} (\bibinfo{year}{1998}).

\bibitem[{\citenamefont{Bordone et~al.}(2004)\citenamefont{Bordone, Bertoni,
  Rosini, Reggiani, and Jacoboni}}]{Bertoni}
\bibinfo{author}{\bibfnamefont{P.}~\bibnamefont{Bordone}},
  \bibinfo{author}{\bibfnamefont{A.}~\bibnamefont{Bertoni}},
  \bibinfo{author}{\bibfnamefont{M.}~\bibnamefont{Rosini}},
  \bibinfo{author}{\bibfnamefont{S.}~\bibnamefont{Reggiani}}, \bibnamefont{and}
  \bibinfo{author}{\bibfnamefont{C.}~\bibnamefont{Jacoboni}},
  \bibinfo{journal}{Semicond. Sci. Technol.} \textbf{\bibinfo{volume}{19}}
  (\bibinfo{year}{2004}).

\bibitem[{\citenamefont{Kraus}(1971)}]{kraus}
\bibinfo{author}{\bibfnamefont{K.}~\bibnamefont{Kraus}},
  \bibinfo{journal}{Annals of Physics} \textbf{\bibinfo{volume}{64}},
  \bibinfo{pages}{311} (\bibinfo{year}{1971}).

\end{thebibliography}
\end{document}